\documentclass[dvips,12pt]{article}
\def \vc #1  {{\bf #1}}

\newcommand{\lr}{\left(}
\newcommand{\rr}{\right)}
\newcommand{\lfc}{\left\{}
\newcommand{\rtc}{\right\}}
\newcommand{\lta}{\left\langle}
\newcommand{\rta}{\right\rangle}

\newcommand{\eps}{\varepsilon}

\newcommand{\sprd}[2]
{\left\langle#1\,,#2\right\rangle\,}
\newcommand{\prt}{\partial}
\def \moh {{-1/2}}
\def \as {\,^*}
\def \elmg {electromagnetic }
\def \om {\omega}
\def \wd {\wedge}

\def \va {\vc A }
\def \vac {\bar{\vc A }}

\def \vcx {\dot{\vc x }}
\def \dx {\delta\vc x }
\def \vca {\dot{\vc A }}
\def \vcca {\dot{\bar{\vc A }}}
\newcommand{\Ds}[1]{{D#1\over ds}}

\newcommand{\lacm}[1]{{\left\{#1\right.}}
\newcommand{\racm}[1]{{\left.#1\right\}}}
\newcommand{\cc}[1]{\bar{#1}}
\newcommand{\half}{\frac{1}{2}}

\def \vce {{\vc e }}
\def \cdt {\hspace{0.03em}\cdot\hspace{0.03em}}
\def \fdt {\hspace{0.36em}}
\def \dtx {\dot{x}}

\newcommand{\gamm}[3]{\,\gamma_{#1#2\cdt }^{\fdt\fdt#3}\,}
\newcommand{\con}[2]{\omega_{#1\cdt}^{\fdt#2}}
\def \ccn {\mbox{C.C. }}
\begin{document}
\markboth{}{Electromagnetic field with constraints and Papapetrou
equation}

\title{Electromagnetic field with constraints and Papapetrou equation}

\author{Turakulov Z. Ya.\\
\footnotesize National University of  Uzbekistan, Vuzgorodok, Tashkent
700174 Uzbekistan\\ \footnotesize Astronomy Institute, and Institute
of Nuclear Physics, Ulughbek, Tashkent
702132 Uzbekistan\\ \footnotesize Tashkent, Uzbekistan\\ Muminov A. T.
\\ \footnotesize Institute of Applied Physics affiliated to the
National University of  Uzbekistan\\ \footnotesize \\
\footnotesize{at-muminov@bcc.com.uz}}

\maketitle

\begin{abstract}
It is shown that geometric optical description of \elmg wave with
account of its polarization in curved space-time can be obtained
straightforwardly from the classical variational principle for \elmg
field. For this end the entire functional space of \elmg fields must
be reduced to its subspace of locally plane monochromatic waves.
We have formulated the constraints under which the entire functional
space of \elmg fields reduces to its subspace of locally plane
monochromatic waves. These constraints introduce variables of another
kind which specify a field of local frames associated to the wave and
contain some congruence of null-curves. The Lagrangian for constrained
\elmg field contains variables of two kinds, namely, a congruence of
null-curves and the field itself. This yields two kinds of
Euler-Lagrange equations. Equations of first kind are trivial due to
the constraints imposed. Variation of the curves yields the Papapetrou
equations for a classical massless particle with helicity 1.
\end{abstract}
\section{Introduction}

~~~Quantum mechanics provides exhaustive description of motion of a
particle in limited scales, when typical length of run is comparable
with the wavelength, normally, in atomic and sub-atomic ones. When
considering motion of a particle in scales which are apparently
non-comparable with typical wavelength, that exhaustive quantum
mechanical picture becomes less convenient, and to use it, one passes
to asymptotical behavior of incident and scattered waves. In
astrophysical scales, particularly, when studying deflection of light
in gravitational field, classical mechanics is evidently more
convenient and one prefers to consider photon as a classical massless
particle drawing a null-geodesic in the space-time. This approach to
light propagation in curved space-time is quite satisfactory while
photon is considered as a scalar particle. If, however, its
polarization becomes important the question arises, how to include it
into the classical mechanical description.

If the wavelength is big enough, i.e., photon momentum is relatively
small, the commonplace classical mechanical description becomes
incomplete. The point is that under some conditions spin of the
particle becomes comparable with some components of its orbital
momentum, which usually are regarded as zero. This may happen, for
example, when describing a light beam incident to a Schwartzschild
black hole. Due to the classical mechanical considerations each photon
of the beam has zero longitudinal component of orbital momentum.
However, its spin is also longitudinal and, hence, contributes the
total longitudinal component of angular momentum. Since, on one hand,
spin is always collinear to momentum of photon, and, on the other
hand, gravitational deflection of the beam changes the momentum, this
description contradicts the conservation law of total angular
momentum. It is clear that in order to have the correct picture one
should take spin of photon into account such a way that the total
angular momentum conserves. If it is done the sum of longitudinal
components of orbital and inner momenta is constant, thus, the earlier
is not zero after deflecting. Therefore, if photon momentum is small
enough, this effect can change the shape of scattered beam. Thus,
there exist situations when the well-known corpuscular theory does
not work.

While in case of spinless particle one can make a choice between the
usual (corpuscular) and wave mechanics due to scales under
consideration, in case of particle with helicity the earlier does not
work, so, the only possibility is to use the latter. The latter
provides description in terms of special functions, say, Legendre
polynomials, which are convenient while their powers are not very
high. However, in some real situations powers of the polynomials are
of the order of astronomical distances measured in wavelength units.
In these situations corpuscular description in which spin of the
particle is taken into account properly, would be much more
convenient. The goal of the present work is construct a model of
photon with given momentum and helicity $\pm1$ in a curved space-time.
\section{Statement of the problem}

~~~The desired model is assumed to provide certain world line of a
massless particle which has helicity 1 and, at the same time,
description of \elmg field which everywhere draws a locally plane
wave. The main difficulty is that spin of the particle is quantized,
thus, the desired construction must contain both classical and quantum
degrees of freedom. The notion of classical particle with quantum spin
seems to be one of the simplest systems of mixed classical-quantum
nature, and attempts to built correct theory of this object are
lasting for decades \cite{frszk}. Our task is somewhat wider, because
we try not only to obtain equation of the particle world line, but
also a locally plane wave, in other words, to describe propagation of
circularly polarized photon in terms of both geometric and wave
optics.

An attempt to build such a model was made in the works \cite{zr1,zr2}
due to the problem of light propagation in Schwartzschild
space-time. Since \elmg field presents in the construction a new
question arises, how to combine the field in the space-time and
helicity which must be attached to the unknown world line. In the work
\cite{zr1} \elmg field was removed by a vector field which was defined
on the world lines. This substitution made it possible to combine wave
and world line under assumption that if the lines are found properly
then the fields attached to them constitute the entire \elmg field in the
space-time.

In the present work we revise this approach. Instead of specifying a
functional space of curves and attaching a vector field to each curve,
we assume that the same result could be obtained from the pure \elmg
Lagrangian. The main idea of this work is that, after all, both
geometric and wave optics should follow from pure \elmg theory,
therefore, the model should be built in the framework of the theory of
\elmg field. To do it, we start with the well-known form of Lagrangian of
\elmg field and restrict the functional space of the field variables
with its subspace of the fields behaving as locally plane waves.
Since restrictions of this sort are known as constraints we assume
that putting relevant constraints on the Lagrangian leads to special
Euler-Lagrange equations for the waves in question, and the desired
description containing both geometric and wave optics follows from it.
\section{Locally plane wave and associated orthonormal frame}

The notion of plane wave in space-time differs from that in space
where the wave vector is orthogonal to the hyperplanes, cannot be
tangent to them. In space-time the wave vector is orthogonal to the
hyperplanes and, at the same time tangent to them. To see this
consider flat space-time and Cartesian coordinates $\{t,x,y,z\}$ in it
which are chosen such a way that the wave has phase $\phi=\om(t-z)$.
The phase takes constant values on luminal hyperplanes $t=z$ and the
wave vector $\vce_-= 2^\moh(\prt_t+\prt_z)$ tangent to the
hyperplanes:\begin{equation}\label{tangent}\vce_-\circ\phi=0.
\end{equation}At the same time the wave vector is orthogonal to the
hyperplanes because, due to pseudo-Euclidean metric any null-vector is
orthogonal to itself. The wave propagates along this vector, therefore
$\vce_-$ must be identified with the vector of velocity and plays the
role of velocity of photon in corpuscular model. Now we use these two
objects to construct an orthonormal frame associated with the wave.

By construction, there exists an isotropic vector $\vce_-$ tangent
everywhere to the surfaces of constant phase, and, therefore,
orthogonal to them. We introduce one more isotropic vector $\vce_+$
whose direction is arbitrary, requiring only that its scalar product
with the vector $\vce_-$ is equal to one. As it is done we can
introduce two unit space-like vectors $\vce_\alpha$ which are
orthogonal to each other and to $\vce_\pm$. The four vectors defined
this way constitute a local orthonormal frame.

In Minkowski space-time and Cartesian coordinates considered above the
space-like vectors $\vce_\alpha$, $\alpha=1,2$ are defined as follows:
$$\vce_1=\prt_x,\,\,\vce_2=\prt_y.$$They are also tangent to the wave
fronts and orthogonal to them. These vectors are used for specifying
polarization of the wave. The four vectors $\vce_\alpha,\,\vce_\pm$
where coordinates are chosen such that $\vce_+=2^\moh(\prt_t-\prt_z)$
form an orthonormal frame with metric\begin{eqnarray}\label{onv}
\hspace{0.2cm}<\vce_+,\,\vce_+>=
<\vce_-,\,\vce_->=<\vce_\pm,\vce_\alpha>=0,\\
\nonumber <\vce_-,\,\vce_+>=1,\, <\vce_\alpha,\vce_\beta>=
-\delta_{\alpha\beta},\hspace{0.2cm}\alpha,\beta=1,2.\end{eqnarray}
The frame of 1-forms dual to it, is\begin{eqnarray*}
\theta^-=dt+dz,\,\,\theta^+=dt-dz\\ \theta^1=dx,\,\,
\theta^2=dy\end{eqnarray*}Now, let us pass to a curved space-time and
account main features of an \elmg wave which can be called locally
plane and monochromatic. We start with a wave which possesses locally
a scalar function $\phi$ called phase. Gradient of this function is an
isotropic 1-form, and its (hyper-)surfaces of level are wave fronts,
i.e., it is possible to introduce local Cartesian coordinates in which
the wave can be represented locally the way just discussed. This is
possible under some special condition, due to which the wavelength is
much less than typical scales specified by the space-time curvature.
Hereafter we assume that this condition is satisfied. The field of
orthonormal frames associated with the wave can be constructed
similarly.

As this is done it remains to fix one detail. The point is that the
metric of this frame given by the equations (\ref{onv}) remains
unchanged if one of isotropic vectors is multiplied by an arbitrary
factor and another one is divided by it. Employing this operation we
can fix action of the vector $\vce_-$ on the phase:\begin{equation}
\label{omega}\vce_+\circ\phi=\om,\quad\om=const\end{equation} while,
by construction, the equation (\ref{tangent}) remains in force. Now,
as the orthonormal vector frame associated with the wave
$\{\vce_\pm,\,\vce_\alpha\}$ is defined, one can find out the
orthonormal covector frame $\{\theta^\pm,\,\theta^\alpha\}$ dual to
it and the connection 1-form $\con{a}{b}\equiv\gamm{c}{a}{b}\theta^c$
for the frame. Hereafter we assume that this is done, and pass to
considering constraints whose imposed on \elmg field leaves only
locally plane waves.\section{Electromagnetic field with constraints}

~~~If the field in question draws everywhere a locally plane wave
there exists a congruence of null-curves $\{x^i(s)\}$ which are
integral lines of the vector $\vce_-$. Since this vector is
identified with the velocity the wave vector is collinear to it,
and potential of the field $\alpha$ has no $\pm$ components, so,
we have a constraint given by\begin{equation}\label{constr1}
\alpha=A_\beta\,\theta^\beta,\,\beta=1,2.\end{equation}The vectors
$\vce_1,\,\vce_2$ are chosen to specify polarization of the wave,
hence, their span does locally the instant (two-dimensional
space-like) wave front. Therefore we can neglect changes of the
potential in their directions. This constraint can be written as
\begin{equation}\label{constr2}\vce_1\circ\alpha=\vce_2\circ\alpha=0
\end{equation}where the operator $\vce_\beta\circ$ stands for
differentiation along the vector $\vce_\beta$.

By analogy with plane waves in Cartesian coordinates the amplitudes
can be represented in the form\begin{equation}\label{constr3}
A_\beta=a_\beta\,e^{i\phi}\end{equation}where the contants $a_\beta$
are complex numbers chosen such a way that the wave has left or right
circular polarization and the function $\phi$, specifies the phase of
the wave. In these denotions the constraint (\ref{constr2}) coincides
with the equation (\ref{tangent}). Finally, all the constraints
imposed above (\ref{tangent}, \ref{omega}-\ref{iomega}) reduce to
the following: the 1-form of field potential has only one non-zero
derivative \begin{equation}\label{iomega}\vce_+\circ\alpha=i\om\,
\alpha,\,\vce_-\circ\alpha=\vce_\beta\circ\alpha=0\end{equation}where
$\phi$ and $\om$ do not depend on the parameter $s$.\section{Action
principle for constrained fields}

~~~The action functional for \elmg field is given by the well-known
integral\begin{equation}\label{act1}
{\cal A}=\int d\alpha\,\wd\as d\alpha=\int<d\alpha,\,d\alpha>\eps,
\end{equation}where $\alpha$ is 1-form of potential of the field,
$<,>$ stands for scalar product and $\eps$ denotes the unit
4-form: $\eps\equiv\eps_{ijkl}/4!\hspace{0.5em}\theta^i\wd
\theta^j\wd \theta^k\wd \theta^l$ that corresponds to four-dimensional
integration in the space-time. Straightforward computation of
variation of the action (\ref{act1}) yields Maxwell equations,
valid for all possible shapes of \elmg field. Our goal is to
restrict the functional space of the field with subspace of fields
which draw everywhere locally plane waves by imposing constraints
((\ref{tangent}, \ref{omega}-\ref{iomega})).

For this end we expand the quadric form $<d\alpha,d\alpha>$ in the
frame $\{\theta\}$ as follows:\begin{eqnarray*}<d\alpha,d\alpha>=
\lr\vce_a\circ A_b\rr\lr\vce_c\circ A_d
\rr\sprd{\theta^a\wd\theta^b}{\theta^c\wd\theta^d}\\
=\sum_{a,\,c=\pm}\lr\vce_a\circ A_b\rr\lr\vce_c\circ
A_d\rr\sprd{\theta^a\wd\theta^b}{\theta^c\wd\theta^d}\\
=\sum_{a,\,c=\pm}\lr\vce_a\circ A_b\rr\lr\vce_c\circ
A_d\rr\sprd{\theta^a}{\theta^c}\sprd{\theta^b}{\theta^d}\\
=\sum_{a,\,c=\pm}\sprd{\vce_a\circ\va}{\vce_c\circ\va}
\sprd{\theta^a}{\theta^c},\\ \mbox{where}\quad\va=A^\beta\,
\vce_\beta.\end{eqnarray*}Though most of terms of the expansion are
zero, some of them have non-zero variations. Substituting the
constrained Lagrangian into the action integral (\ref{act1}) yields:
\begin{eqnarray}{\cal A}=\half
\int\lfc\sprd{\vce_+\circ\cc{\va}}{\vce_+\circ\va}
\sprd{\theta^+}{\theta^+}+\sprd{\vce_-\circ\cc{\va}}{\vce_-\circ\va}
\sprd{\theta^-}{\theta^-}+\nonumber\right.\\ \left.
+\lr\sprd{\vce_-\circ\cc{\va}}{\vce_+\circ\va}+\ccn\rr
\sprd{\theta^-}{\theta^+}\rtc\eps.\label{act2}\end{eqnarray}where
we take into account the fact that for convenience we use complex
valued field components. As usual, the components $\va$ and their
complex conjugates $\cc{\va}$ are regarded as independent
variables. Due to the constraints ((\ref{tangent},
\ref{omega}-\ref{iomega})) the Lagrangian under integral
(\ref{act2}) can be transformed as follows:
\begin{eqnarray*}\lfc\sprd{\vce_+\circ\cc{\va}}{\vce_+\circ\va}
\sprd{\theta^+}{\theta^+}+\sprd{\vce_-\circ\cc{\va}}{\vce_-\circ\va}
\sprd{\theta^-}{\theta^-}+\nonumber\right.\\ \left.
+\lr\sprd{\vce_-\circ\cc{\va}}{\vce_+\circ\va}+\ccn\rr
\sprd{\theta^-}{\theta^+}\rtc=
\om^2\sprd{\cc{\va}}{\va}\sprd{\theta^+}{\theta^+}
+i\om\sprd{\va}{\vcca}+\ccn.
\end{eqnarray*}The third term can be ignored because, as will be shown
below, action of the vector $\vce_-$ annulates the field,
consequently, the expression
$\sprd{\vce_-\circ\cc{\va}}{\vce_-\circ\va}$ is product of two
zero factors, hence, both the third term itself and its variation
are identically zero. Though, due to the constraint
(\ref{tangent}) the term $\vcca$ (and $\vca$) is equal to zero, we
do not ignore it because its variation plays important role in the
action principle. Thus, finally, the action integral has the
form\begin{equation}{\cal A}=\int{\cal L}\,\eps,\quad{\cal L}=
\half\,\om\lfc\om\sprd{\va}{\vac}\sprd{\vcx}{\vcx}+\lr
i\sprd{\va}{\vcca}+ \ccn\rr\rtc.\label{act3}\end{equation}
\section{Helicity of constrained fields}

~~~The action integral is evidently invariant under rotations of the
frame in the plane formed by the vectors $\vce_1$ and $\vce_2$.
This invariance yields some conservation law due to the Noether
theorem. To find the law we consider the change of the form of the
action integral under rotations of the local frames specified by
an infinitesimal matrix $\delta\eta^a_{\,b}(s)$ defined as a
function of the parameter $s$ on each curve. Rotation of the frame
changes components of the vectors $\va,$ $\vac$ but does not change
the vectors. Thus, the first term of the Lagrangian (\ref{act3})
containing scalar product $<\va,\vac>$ does not contribute variation.
At the same time the rotation changes derivative of $\vca$:
\begin{eqnarray*}\delta\vca^a=\delta\left(\frac{dA^a}{ds}+
\gamm{b}{c}{a}\dtx^b A_c\right)=\left(\frac{D\delta A^a}{ds}\right)+
\delta\con{c}{a}(\vcx)\,A^c\vce_a,\\
\delta\con{c}{a}=D_b(\delta\eta_{c\cdt}^{\fdt a})\,\theta^b
\end{eqnarray*}where the covariant derivative
$D_b(\delta\eta_{c\cdt}^{\fdt a})$ is exactly variation of the
connection 1-form.

The first term in the variation of $\vca$ does not contribute
variation of the action due to field equations. By the result,
variation of the action is\begin{eqnarray*}\int\delta{{\cal L}}\eps=
\int\lfc i\om\sprd{\va}{\Ds{\delta{\eta}_{a\cdt}^{\fdt b}}\bar
A^a\,\vce_b)}+ \ccn\rtc\eps=\\=\int\lfc i\om\lr
\delta\eta_{ab}\rr^\bullet\bar A^aA^b+\ccn\rtc\eps=\int
d[\ldots]+\int2\omega\delta\eta_{ab}\Ds{}\lr\frac{\bar A^aA^b-\bar
A^bA^a}{2i}\rr\eps.\end{eqnarray*}Finally, we have:$$\Ds{S_{ab}}=0.$$
So, due to the Noether theorem we obtain conserved spin current with
single non-zero component\begin{equation}\label{spincurr}J^-_{ab}=
J_{+ab}=2S_{ab},\end{equation}where$$S^{ab}=
\frac{\om}{2i}\left(\cc{A}^aA^b-\cc{A}^bA^a\right)$$is the spin tensor
of the wave. Due to the constraint (\ref{constr3}) it is zero for
linearly polarized waves, and for circularly polarized waves has
single non-zero element$$S^{12}=\pm\om|a^1a^2|,$$whose sign depends
only on helicity. The only consequence of this result we need is that
the spin has single non-zero component, and as for its magnitude, we
accept its quantum value 1 in dimensionless units. Note that the spin
is always pointed along the vector of velocity, so no special equation
is needed for it. This fact provides implementation of the Tulczyjew
constraint which requires that conversion of the particle momentum
with its spin is zero \cite{tul}.\section{Field equations}

~~~Now we return to the action functional (\ref{act3}) and compute its
variations only under small variations of the field which has two
components $A_1$ and $A_2$. Variation of the first term in the
constrained Lagrangian is identacally zero because it contains the
factor $\sprd{\theta^+}{\theta^+}$ which is not varied, therefore we
ignore it. Variation of the rest part of the action is
\begin{eqnarray*}\delta{\cal A}=
\delta\int\lfc\lta\vce_+\circ\va,\vce_-\circ\vac\rta+\ccn\rtc\eps=\\
\int\left[\lta\delta\lr\vce_+\circ\va\rr,\vce_-\circ\vac\rta+
\lta\vce_+\circ\va,\delta\lr\vce_-\circ\vac\rr\rta+\ccn\right]\eps=\\
\int\left[\lta\vce_+\circ\delta\va,\vce_-\circ\vac\rta+
\lta\vce_+\circ\va,\vce_-\circ\delta\vac\rta+\ccn\right]\eps
\end{eqnarray*}where action of vectors $\vce_\pm $ is considered as
differentiation along the vectors. The next step is to extract total
derivatives:\begin{eqnarray*}\delta{\cal A}=
\int\left[\vce_+\circ\lta\delta\va,\vce_-\circ\vac\rta+
\vce_-\circ\lta\vce_+\circ\va,\delta\vac\rta+\ccn\right]\eps+\\
-\int\lta\delta\vac,\hspace{0.2cm}\vce_+\circ\lr\vce_-\circ\va\rr
+\vce_-\circ\lr\vce_+\circ\va\rr\rta\eps\\
-\int\lta\delta\va,\hspace{0.2cm}\vce_+\circ\lr\vce_-\circ\vac\rr
+\vce_-\circ\lr\vce_+\circ\vac\rr\rta\eps.\end{eqnarray*}Note that due
to the metric of the null-frame (\ref{onv}) combinations like
$\vce_+\circ V_-$ are parts of divegence of a vector $\bf V$, and if
the vector has only one component $`-'$ this expression coincides with
its divergence. In particular, the combination
$\vce_\lacm{+}\circ\lr\vce_\racm{-}\circ f\rr$ is exactly
Dalembert operator applied to a function $f$ which is constant on the
wave fronts. Consequently, the first term in the right-hand side of
the equation above is exactly a divergence, hence the integral can be
taken by parts and variation of the action integral becomes:
\begin{eqnarray*}=\int d[\ldots\,]-\int\lfc\lta\delta\vac,
\hspace{0.2cm}
\vce_\lacm{+}\circ\lr\vce_\racm{-}\circ\va\rr\rta+\ccn\rtc\eps
\end{eqnarray*}where the first term reduces to a surface integral and
fugure brackets at subscripts mean symmetrization. As usual, the
surface integral vanishes at infinity and all the rest reduces to the
following Euler-Lagrange equation:$$
\vce_\lacm{+}\circ\lr\vce_\racm{-}\circ\va\rr=0.$$Evidently, this
covariant equation reduces to Dalembert equation in local Cartesian
coordinates provided that the curves $x(s)$ are locally null straight
lines. In fact, any vector whose covariant derivative along the curve
is zero:\begin{equation}\label{eq4a}\vce_-\circ\va=0,\end{equation}
and so for the complex conjugate, satisfies this equation. The
constrained fields satisfy this equation due to the equations
(\ref{tangent}-\ref{constr3}), consequently, this part of the entire
variation of the action integral is zero due to the constraints.
Though the field equation leaves $\om$ an arbitrary function of the
phase we restrict our analysis with monochromatic waves for which this
value is constant.\section{Papapetrou equations}

~~~It remains to consider the second part of variation of the action
integral, produced by variation of local frames under fixed field
variables. Since the local frames are defined as co-moving frames
on the congruence of null-curves, it is possible to introduce
variation of the congruence and derive variation of the frames
from it. Small change of shape of a curve $x(s)$ causes small
change of the tangent vector without change of its length.
Consequently, variation of the vector $\vce_-$ is orthogonal to it
and to the complementary null-vector $\vce_+$, hence, belongs to the
span of the two polarization vectors $\vce_\beta$. Since the vector
$\vce_+$ is also orthogonal to the span, it suffers no change.
Therefore, variations of both $\vce_+$ and the corresponding covector
$\theta^-$ are zero. Thus, only variations of the vector $\vce_-$ and
the corresponding 1-form $\theta^+$ contribute variation of the action
integral (\ref{act3}).

The first term in the Lagrangian (\ref{act3}) contains the factor
$\sprd{\theta^+}{\theta^+}=\sprd{\vce_-}{\vce_-}\equiv0$, therefore
its contribution is predetermined only by variation of the vector
$\vce_-=\vcx$. Consider variation of four-dimensional integral
$$\frac{\om^2}{2}\int|\va|^2\sprd{\vcx}{\vcx}\eps$$ with respect to
variation of the curves. Variation of the factor $\sprd{\vcx}{\vcx}$
is well-known from the variation principle for geodesics \cite{pi}.
Here we can use the fact that variation of the integrand reduces to
the scalar product of the vector of variation of the curve $\dx(s)$
and the covariant acceleration $\Ds{\vcx}$:\begin{equation}
\label{accel}\delta\int\frac{\om^2}{2}|\va|^2\sprd{\vcx}{\vcx}\eps=
\om^2\int|\va|^2\sprd{\dx(s)}{\Ds{\vcx}}.\end{equation}

The second term in the Lagrangian (\ref{act3}) has single zero factor
$\vcca$ in the scalar product, consequently, non-zero contribution to
the variation of the action integral appears only when varying this
factor. The zero factor to be varied is $\vcca$:$$\vcca=
\left(\frac{d\bar{A}^a}{ds}+\gamm{b}{c}{a}\dtx^b\bar{A}^c\right)
\vce_a.$$Variation of the covariant derivative contains derivative on
$s$ and the term containing the connection $\gamm{b}{c}{a}$. Variation
of the first of them does not contrtibute variation of the action
integral because neither polarization vectors $\vce_\alpha$ nor
components of the field change under varying the congruence of curves.
The only term suffering some change is connection $\gamm{b}{c}{a}$.
Therefore we can write down contribution of this term as follows:$$
\half\delta\int\om\lr i\sprd{\va}{\vcca}+\ccn\rr\eps=
-\half\int\lr iA^b\delta\gamm{b}{c}{a}\bar{A}^c+\ccn\rr\eps$$where the
sign minus appears due to the lower index at the field component.
Thus, the next task is to find variation of the connection
$\delta\gamm{b}{c}{a}$.

It is easier to find variation of the 1-form $\con{a}{b}$ because the
variation is to be taken in a fixed point under changing the field of
frames which is dragged by the vector $\dx$. This variation is, by
definition, Lie derivative of the connection 1-form with respect to
this vector. So, to find variation of the connection 1-form it
suffices to take its Lie derivative with respect to the vector $\dx$.
Lie derivative of a 1-form $\lambda$ with respect to a vector $\vec v$
is \cite{sz}$$\pounds{}_{\vec v }\lambda=
d\lambda(\vec v)+d(\lambda(\vec v)).$$Unlike ordinary 1-form the form of
connection has components referred to local frames. Since the frames
are built on the vector of velocity on the congruence its variation
causes some infinitesimal rotation of the frames. Denote the
corresponding matrix of rotation $\eta_a{}^c$. This rotation
transforms components of the connection 1-form and must be taken into
account. To do it it is necessary to obtain explicit form of this
matrix from the dragging vector $\dx$.

Consider a point in the space-time a curve passing through it and the
local frame and small variation of congruence of curves given by small
vector $\dx$ which drags the congruence. This dragging replaces the
curve passed through this point with the curve dragged by the vector
and the local frame built in this point is also to be replaced by the
frame dragged by this vector from a neighboring point. Since, on one
hand both the frames are orthonormal variation of the frames is small
rotation. On the other hand, since this rotation is specified by
dragging orthonormal frame from a neighboring point, this
transformation is given by the connection itself, in other words the
matrix of rotation $\eta_a{}^c$ is exactly the value of the form of
connection on the dragging vector: $\eta_a{}^b=\con{a}{b}(\dx)$.

Thus, Lie derivative of the connection 1-form with respect to the
vector $\dx$ is$$\pounds{}_{\dx}\con{a}{b}=d\con{a}{b}(\dx)+
d(\con{a}{b}(\dx))-\eta_a{}^c\con{c}{b}+\eta_c{}^b\con{a}{c}.$$
Subsituting the matrix of rotation we obtain the desirted Lie
derivative:\begin{eqnarray*}\pounds{}_{\dx}\con{a}{b}
=d\con{a}{b}(\dx)-\con{a}{c}(\dx)\con{c}{b}+\con{c}{b}(\dx)\con{a}{c}+
d(\con{a}{b}(\dx))=\\ \\(d\con{a}{b}(\dx)+\con{c}{b}\wd\con{a}{c})
(\dx)+d(\con{a}{b}(\dx))=\Omega_a{}^b(\dx)+d(\con{a}{b}(\dx))
\end{eqnarray*}where we have obtaind the curvature 2-form
$\Omega_a{}^b\equiv R_{cda}{}^b\theta^c\wd\theta^d$. Substituting now
this into variation of $\vcca$ gives:\begin{eqnarray*}\delta\vcca=
R_{dbc}{}^a\delta x^d\dot x^b\bar A^c\vce_a+
\dot x^d\prt_d(\gamm{b}{c}{a}\delta x^b\bar A^c)\vce_a=\\
R_{dbc}{}^a\delta x^d\dot x^b\bar A^c\vce_a+
(\gamm{b}{c}{a}\delta x^b\bar A^c)^\bullet\vce_a.\end{eqnarray*}
Variation of $\vca$ is similar, so after composing the total variation
of the second term in the action integral we obtain the two terms. One
is total derivative of $\om(A_a\gamm{b}{c}{a}\delta x^b\bar A^c)$ on
$s$, which vanishes on the endpoints of the curves. Thus, the whole of
variation of this part of Lagrangian is given by another term which is
$$\frac{i\om}{2} R_{dbc}{}^a\delta x^d\dot x^b(\bar A^cA_a-\bar A_aA^c)=
-R_{dbc}{}^a\delta x^d\dot x^bS^c{}_a$$where we introduce spin by
its only component $S^1{}_2$. The Euler-Lagrange equation for the
curves $x(s)$ coincides with Papapetrou equation:\begin{equation}
\label{papa}\om^2|\va|^2\Ds{\dtx^a}=R_{db\cdt c}^{\fdt\fdt a}\,
\dot x^cS^{db}.\end{equation}\section{Conclusion}

~~~Geometric-optical description of \elmg wave in curved space-time,
with account of its polarization is obtained straightforwardly from
the classical variational principle for \elmg field. For this end the
entire functional space of \elmg fields is reduced to its subspace of
locally plane monochromatic waves. Therefore, first of all, the notion
of locally plane monochromatic wave in curved space-time should be
defined. It turns out that waves of this sort exist provided that
their wavelengths are small compared with scales under consideration.
Assuming this, we have formulated the constraints under which the
entire functional space of \elmg fields reduces to its subspace of
locally plane monochromatic waves and imposed these constraints.
These constraints not only reduce field variables but also introduce
variables of another kind which specify a field of local frames
associated to the wave and contain some congruence of null-curves
$x^i(s)$. These curves become the main object in the construction
because it specifies the field of local frames and the field variables
$\vca$ and $\vcca$ are referred to this frame.

Returning to the action principle for the constrained \elmg field we
have Lagrangian (\ref{act3}) which contains variables of two kinds,
namely, a congruence of curves $x^i(s)$ and the field itself and have
two kinds of Euler-Lagrange equations. Equations of first kind reduce
to local Dalembert equation for the field components $\vca$ and
$\vcca$ which are trivial due to the constraints imposed. Variation of
the curves yields all the rest equations which contain the main result
of this investigation. It turns out that the Euler-Lagrange equations
they yield are exactly the Papapetrou equations for a classical
massless particle with helicity 1. This equation determines the shape
of the 0-curves which, by construction, can be considered as world
lines of photons with the same wavelengths and helicities. They
apparently differ from null-geodesics and, thereby manifest influence
of spin-gravitational interaction on propagation of \elmg waves in
gravitational fields. Effect of this interaction is proportianal to
the wavelength \cite{zr2}, therefore this fact can be observed in
radioastronomy.\end{document}